\documentstyle[preprint,aps,epsf,prb]{revtex}
\draft
\begin{document}
%
\title{Critical behavior of the restricted primitive model revisited} 
\author{J.-M. Caillol, D. Levesque and J.-J. Weis}
\address{Laboratoire de Physique Th{\'e}orique, UMR 8627,
  B{\^a}t. 210, Universit{\'e} de Paris-Sud \\ 91405 Orsay Cedex,
  France}
\date{\today}
\maketitle
\begin{abstract}
Reassessment of the critical temperature and density of the restricted
primitive model of an ionic fluid by Monte Carlo
simulations performed for system sizes with linear dimension up to
$L/\sigma=34$ and sampling of $\sim 10^9$ trial moves leads to
$T^*_c=0.04917 \pm 0.00002$ and $\rho_c^* =0.080 \pm 0.005$. Finite size
scaling analysis based in the Bruce-Wilding 
procedure gives critical exponents in
agreement with those of the 3d Ising universality class. An analysis
similar to that proposed by Orkoulas {\em et al} [Phys.\ Rev.\ E {\bf
63}, 051507 (2001)], not relying on an  {\em a
priori} knowledge of the universality class, leads to an unaccurate
estimate of $T_c^*$ and to unexpected behavior of the specific heat and
value of the critical exponent ratio $\gamma/\nu$.
\end{abstract}
\newpage
\section{Introduction}
Despite endeavor over more than a decade to elucidate the
nature of the critical behavior of the restricted primitive model (RPM)
for ionic fluids, prototype of a system governed by long range Coulomb
interactions, no unassailable answer to this question has yet been
provided by theory, experiment or computer simulation \cite{fish,stell,fisher:1996,stell:1996,stell:1999,wein}. The long
range character of the interaction would suggest classical (mean field)
behavior, whereas the well known  screening of the interactions
pleads in favor of an Ising-type criticality typical of systems with short
range interactions. In contrast to the latter case a rigorous
renormalization group (RG) treatment allowing to decide in favor of one or
the other universality class is, for the moment, still unavailable due
to the lack of a satisfactory mean field starting point for RG analysis \cite{more,ciach}.

On the experimental side \cite{wein} an indisputable interpretation of
criticality in ionic systems
and assessment of the role played by the Coulomb interaction 
is somewhat hampered by the possible interplay of the Coulomb
interaction with other forces driving phase separation (as, for
instance, solvophobic effects), uncertainties of measurement close to
the critical point or choice of appropriate order parameter to analyse
the results. It seems however well established now that for many
experimental systems apparent mean field behavior applies with sharp
crossover (much sharper than in non-ionic fluids) to Ising criticality
close to the critical temperature \cite{wein}.

Computer simulations can isolate the effect of the Coulomb interactions
but are plagued with their own difficulties when approaching the
critical region, in particular by the limited system sizes that are
currently accessible for off-lattice systems. Finite size corrections to
the scaling behavior may therefore be important and thwart extrapolation
to the thermodynamic limit. In addition, the fact that the RPM, as well
as most realistic continuum models, lack symmetries, that are
present, for instance, in the Ising or lattice-gas models has the consequence that the asymptotic scaling
properties are more complex than for the latter systems. 

Starting with the seminal work of Bruce and Wilding (BW)
\cite{bruce:1992,wil1} simulation results
for the critical behavior of asymmetric fluids have customarily been
analyzed along the lines of the revised scaling theory of Rehr and
Mermin \cite{rehr} in which the two relevant scaling fields $h$ (the
strong ordering 
field) and $\tau$ (the weak thermal field) are assumed to be linear
combinations of deviations from the critical values of the chemical
potential $\mu$ and the inverse temperature $\beta =\displaystyle
\frac{1}{kT}$ (for the definition of the reduced quantities see below)
\begin{equation}
h = \mu^*-\mu_c^*+ r(\beta_c^*-\beta^*)
\label{a1}
\end{equation}
\begin{equation}
\tau = \beta_c^*-\beta^* + s(\mu^*-\mu_c^*)
\label{a2}
\end{equation}
where $s$ and $r$ are system dependent coefficients defined in Ref.\ \onlinecite{rehr}. 
The revised scaling theory assumes  analyticity of  $\mu (T)$ at  the
 critical temperature $T_c$. Although this is the case for
the Ising model and some models with ``hidden'' symmetry, there is  no
compelling reason that, in general, for fluid systems $\mu (T)$  should lack a  singularity as
recognized already by Rehr and Mermin \cite{rehr}, Yang and Yang
 \cite{yang} and emphasized more recently by Fisher and coworkers
 \cite{fish:2000,orkou:2000,orkou2:2001}. The latter authors, by carefully analyzing experimental results for the constant volume
heat capacity 
\begin{eqnarray}
C_V(T)& = &VT {(\partial^2 p/ \partial T^2)}_V
         -NT {(\partial^2 \mu / \partial T^2)}_V \\
      & = & C_p + C_{\mu} 
\label{cv}
\end{eqnarray}
give evidence for a divergence of ${(\partial^2 \mu / \partial T^2)}_V$,
the so called Yang-Yang (Y-Y) anomaly \cite{yang},
in CO${}_2$ and propane when approaching the critical point from below,
e.g.\ along 
the critical isochore  \cite{fish:2000,orkou:2000}.
According to Fisher and Orkoulas, in order to accomodate the Y-Y anomaly, the pressure
should combine with $\beta$ and $\mu$ in Eqs.\ \ref{a1} and \ref{a2}
 \ \cite{fish:2000}.
 This in turn will affect the finite size scaling
(f.s.s.) analysis - at the core of all simulation studies - through
appearance of additional size dependent terms
which may compete with those of the customary description.
One can note, however, that
for a hard core square-well model fluid the strength of the Y-Y anomaly
appears to be quite small \cite{orkou2:2001}.

The present study was undertaken to extend simulation work on the RPM
\cite{cai1,cai2} to
system sizes larger than previously considered, covering the volume range
$(5000-40000)\sigma^3$ (or linear dimension $L=17-34 \sigma$), 
thus providing
a valuable check of the validity of previous extrapolations of the
critical parameters to their thermodynamic limit.
By the same token statistics of the runs performed previously with the
smaller system sizes were considerably increased.  
In addition, these new simulations gave us the opportunity to
investigate the behavior  of the two contributions $C_p$ and $C_{\mu}$
to the
specific heat near its critical point.
The occurrence of a divergent $C_{\mu}$ would call for
a revision of the revised scaling assumptions of Rehr and Mermin
\cite{rehr} as
pointed out by Fisher and Orkoulas \cite{fish:2000}.

The model and a few computational details are given in Sec.\ II and the
results in Sec. III. The conclusions are summarized in Sec.\  IV.

\section{Model and boundary conditions}

In the RPM of an ionic solution $N/2$
particles carrying a charge $+q$ and an equal number of particles 
with charge $-q$ interact via a hard sphere excluded volume and a
Coulomb
interaction, i.e.\
\begin{eqnarray}
\label{pot}
v_{ij}(r)& =&\left\{ \begin{array}{ll}
+\infty  & r \le \sigma \\
\displaystyle\frac{q_i q_j}{D} \frac{1}{r}
& r> \sigma 
\end{array} \right. 
\end{eqnarray}
where
 $\sigma$ is the hard sphere diameter  and $D$ the dielectric constant
 of the solvent assumed to be a dielectric continuum.

In fact a thermodynamic state is  specified by the combination
 $ \displaystyle T^{*}=\frac{kTD\sigma}{q^2}$ defining a reduced temperature (or its
 inverse $\beta^* = 1/T^*$) and a  reduced chemical
 potential $\mu^*= \mu/kT -3 \ln(\Lambda/\sigma$) ($k$ Boltzmann's
 constant, $\Lambda$  the de Broglie thermal wavelength).  
A reduced 
density is defined as $\rho^*
= N\sigma^3/V$ ($N$ total number of ions, $V$ volume).

When hyperspherical boundary conditions  \cite{cai:1993} are used,
as done 
here in accord with our previous work \cite{cai1,cai2}, the
particles are confined to 
the surface $S_3$ of a hypersphere in 4d space.
In this geometry the RPM may be viewed as a system of identical
particles of charge $q$ (bicharges) interacting by the potential \cite{cai:1993}
\begin{equation}
v_{ij}^{el,S_{3}}= \frac{q^2}{R} \cot \psi_{ij}  \  (0 < \psi_{ij} < \pi).
\end{equation}
The distance $r_{ij}$ between two particles  on $S_3$, measured along
the geodesic joining them, is related to the angle $\psi_{ij}$ by
$r_{ij} = R \psi_{ij}$
where $R$ is the radius of the hypersphere.
The Monte Carlo (MC) simulations were performed in the grand canonical (GC) ensemble
using a biasing scheme \cite{cai2,orkou:1994} to enhance the acceptance ratio of the trial
insertion and deletion moves. 
During the simulation runs we recorded, at fixed $\mu$, $T$ and $V$ the
joint distribution $p_L(\rho,u)$ of particle number and energy density
$u=U/V$ which is the basic ingredient for our analysis of the critical
properties. 
Use of histogram reweighting \cite{fer} was made to infer the
distribution at a state ($\beta, \mu$) from the known one at a nearby state
($\beta_0, \mu_0$).

\section{Results}
In this section we intend to reassert, within the mixed-field
f.s.s. approach of Bruce and Wilding \cite{wil1} , our previous
estimates \cite{cai2} of the
critical parameters  taking into account new simulations at volumes
$V/\sigma^3=20000$ and $40000$ and results with increased statistics at 
$V/\sigma^3=5000$ and $10000$.
Briefly stated, in this approach the appropriate scaling operators
conjugate to the scaling fields $h$ and $\tau$ (Eqs \ref{a1} and \ref{a2}) are assumed to
be  
\begin{equation}
\delta {\cal M} = {\cal M} - <{\cal M}>_c
\label{b12}
\end{equation}
\begin{equation}
\delta {\cal E} = {\cal E} - <{\cal E}>_c
\label{b13}
\end{equation}
where
\begin{equation}
{\cal M}\  =\ \frac{\rho -su}{1-sr}
\label{b14}
\end{equation}
\begin{equation}
{\cal E}\ =\  \frac{u-r\rho}{1-sr}
\label{b15}
\end{equation}
and ${<\cal M>}_c$ and ${<\cal E>}_c$ are the values at criticality.

With this postulate the critical behavior of the fluid system can be
mapped on that of the (symmetric) Ising spin system. In particular, the
distribution $p_L(\cal M)$ of the ordering parameter  should be invariant 
under the symmetry transformation $\delta \cal M \to  -\delta \cal M$
along the coexistence curve $h=0$ and similar to that of the 3d Ising magnetization. 
The strategy offered by Bruce and Wilding \cite{wil1} to determine the critical
parameters $T_c$ and $\rho_c$ is to vary $\mu,T$ and $s$ until the
distribution $p_L(\cal M)$ derived from $p_L(\rho,u)$ (measured in the
simulation) through the linear transformations Eqs.\  \ref{b14} and \ref{b15} 
and integration over $\cal E$ matches the  distribution of the 3d
Ising universality class  $p^*_{is}$ known from lattice spin simulations \cite{hilf,tsyp}. 

Due to the finite size of the simulation volumes the critical parameters
so obtained will be shifted with respect to their infinite volume
values. Finite size scaling theory tells us, however, how the apparent
parameters scale with system size $L$. The critical temperature, for
instance, should vary as
\begin{equation}
T^*_c(L)-T^*_c(\infty) \propto L^{-1/\nu -\theta/\nu} 
\label{scalt}
\end{equation}
where allowance has been made for correction to scaling through the
Wegner exponent $\theta$ \cite{wegner:1972}.

The thermodynamic states at which histograms were recorded are
summarized in Table I. All simulation runs, including those at $V/
\sigma^3=5000$ and 10000 are new. The total number of selected
configurations, spaced by 250 MC trial moves, varies between 
$10^8$ and 4 $10^8$ depending on volume (cf.\ Table I)
 and is thus 4-10 times larger than that generated in Ref.\
\onlinecite{cai2}.
By histogram reweighting we estimated, for each volume, an apparent critical
temperature such that the order parameter distribution $p_L({\cal
M})$, normalized to have unit variance, matches
the 3d Ising model universality class.

The matching of $p_L({\cal M})$ and $p_{is}^*({\cal M})$ at $T^*_c(\infty)$
was realized in  Ref.\ \onlinecite{cai2}
using the estimate of $p_{is}^*({\cal M})$ made by 
Hilfer and Wilding \cite{hilf} for the 3d Ising model on cubic lattices
of sizes $20^3$ and $30^3$.
Recently two new estimates of $p_{is}^*({\cal M})$ have been obtained
by Tsypin and Bl\"ote \cite{tsyp} for the 3d Ising model and the spin-1
Blume-Capel model with lattice sizes up to $58^3$.
The two evaluations of $p_{is}^*({\cal M})$ at $T^*_c(\infty)$ differ
notably, especially for the values of the two maxima of
$p_{is}^*({\cal M})$ 
 at ${\cal M} = \pm {\cal M}_{max}$. However,  Tsypin and Bl\"ote \cite{tsyp}
consider the distribution $p_{is}^*({\cal M})$ evaluated for
the Blume-Capel model to be the more reliable since finite size effects appear
to be
smaller for the largest lattice sizes considered in their simulations.
In view of these differences it seemed thus justified, in order to
determine the apparent critical temperatures  $T^*_c(L)$ for the different volumes, to realize the matching 
of $p_L({\cal M})$ using 
$p_{is}^*({\cal M})$ obtained with the Blume-Capel model.
For volume sizes $V/\sigma^3 \le 10000$ the matching procedure is not without
ambiguity due to poor or insufficient sampling of densities smaller or
close to $2/V$.
In order to minimize the bias on  $T_c^*(L)$ introduced by insufficient
sampling of the low densities the matching has been realized, for each
volume, by determining the smallest mean square deviation between
$p_{is}^*({\cal M})$ and $p_L({\cal M},T,\mu ,s)$ by minimizing
\begin{equation}
\chi_L^2 = \int_{-1.5}^{1.5}(p_{is}^*(x) - 
p_L(x,T,\mu,s))^2 dx
\label{xi}
\end{equation}
in the domain of values where $p_L({\cal M})$ seems most reliable
with the constraint that $\mu$ and $s$ are such that
 $p_L(x_{max})$ = $p_L(-x_{max})$. Here $x= \delta {\cal M}/\sqrt{<{\delta {\cal M}^2}>}$

Figure \ref{fig1} shows $\chi_L^2$ as a function of $s$ in the vicinity
of $T_c^*(L)$ for the four volumes considered and   $p_{is}^*({\cal M})$ given
by the Blume-Capel model \cite{tsyp}.
For $V/\sigma^3 = 5000$ there are two equivalent minima at $T_c^* =
0.004934$ for $s=-1.44$ and $s=-1.45$. The latter value has been retained.
A similar minimization has also been realized using $p_{is}^*({\cal M})$
calculated for the 3d Ising model \cite{tsyp}. In this way one obtains two sets of
values of $T_c^*(L)$, plotted in Fig.\ \ref{fig2}, from which $T_c^*(\infty)$ can be determined using 
Eq.\ \ref{scalt} for the extrapolation of the $T_c^*(L)$ as a function of 
$L^{-(\theta +1)/\nu}$. 
These extrapolations lead to the estimates $ 0.04917 \pm
0.00002$ using $p_{is}^*({\cal M})$ derived from the Blume-Capel model
and $0.04916 \pm 0.00002$ using $p_{is}^*({\cal M})$ obtained with the
3d Ising model.
The errors on $T_c^*(L)$ correpond to those on the
localisation of the minimum of $\chi_L^2$.
In the following  $p_{is}^*({\cal M})$ will refer to the universal
distribution obtained from the Blume-Capel model.

Use of this new determination of $p_{is}^*({\cal M})$ leads to an increase of
the critical 
temperature $T_c^*$ by  $\sim 0.5 \%$ with respect to our previous
estimate \cite{cai2}. It is worth noticing that the latter estimate of
$T_c^*$ included volumes $V/\sigma^3 \le 5000$ for which the region of very low
density states ($ \le 2/V$ ) cannot be sampled.

The collapse of $p_L({\cal M})$, obtained by minimizing $\chi_L^2$,
 on the universal distribution
$p_{is}^{*}$  for the different volumes is shown in Fig.\
\ref{fig3} .
At volume $V/\sigma^3=5000$ a mismatch is observed at the lowest values of
${\cal M}$ due, as explained in Ref.\ \onlinecite{cai2}, to inadequate sampling of the low
density configurations at small volume. 
At volumes $V/\sigma^3= 10000 $ and $20000$ the
agreement is excellent. It is less good at the larger volume, especially
for $V/\sigma^3=40000$.
The most plausible explanation for this discrepancy is a statistical effect
due to insufficient sampling of the region of densities comprised
between the high and low density maxima. We attempted to improve the sampling by
using a multicanonical
method \cite{wil1,berg}, which permits enhanced
crossing of the free energy barrier separating the gas and liquid
phases, but did not  observe a sensible reduction of the discrepancy.

From the knowledge of the order parameter distribution we can calculate
the ratio 
\begin{equation}
Q_L  =  \frac{ {{<\delta {\cal M}^2>}_L}^2}{{<\delta {\cal M}^4>}_L}
\label{b22}
\end{equation}
which takes a well-defined universal value $Q^*$ at $T=T_c$ and $L
\rightarrow \infty$ \cite{binder:1981,blote:1995}.
From f.s.s. theory it follows that $Q_L$ can be expanded in the vicinity
of the critical point as \cite{blote:1995}
\begin{eqnarray}
Q_L(\beta^*) & = & Q^* +q_1 (\beta^* -\beta_c^*) L^{1/ \nu} 
+q_2 (\beta^* -\beta_c^*)^2 L^{2/ \nu} \nonumber \\
     & & +q_3 (\beta^* -\beta_c^*)^3
L^{3/ \nu} \mbox{} + \cdots + b_1 L^{y_{i}} + \cdots. 
\label{b25}
\end{eqnarray}
where the last term takes into account contributions from irrelevant
fields and $q_1$, $q_2$, $q_3$ and $b_1$ are non-universal constants.  
For each volume $V$ and $T$ in the vicinity of 
$T_c^*(\infty)$, $Q_L$ has been determined by calculating the moments of
the symmetrized distribution   $p_L({\cal M},T)$, i.e. such that
 $p_L({\cal M}_{max})$ = $p_L(-{\cal M}_{max})$ for an appropriate
choice of $\mu^*$, $s$ having the value corresponding to that which
realizes the matching of $p_L({\cal M})$ at $T_c^*(\infty)$
since, as apparent from Fig.\ \ref{fig1}, $s$ depends weakly on $T$ at
given volume.

A fit of $Q_L(\beta^*$) obtained for the four volumes along the
coexistence curve turned out not to be possible, within the present
precision of data, when $\beta_c^*$, $Q^*$, $q_1$, $q_2$, $q_3$, $b_1$ and the
exponents 
$\nu$, $\theta$ and $y_i$ were all considered as free parameters. In contrast,
when fixing $\beta_c^*$ to the value derived above, $\beta_c^* =
1/0.04917$, and using the value $y_i =-\theta/\nu=-0.84$ a fit better than 1\% is
obtained giving $Q^* \approx 0.63 \pm 0.01$ and $\nu = 0.66 \pm 0.03$.
Conversely, if $Q^*$ is fixed at the universal value of the Ising class
and $\theta=0.53$, all other parameters being left
free, one obtaines $T_c^*=0.04918$ and $\nu = 0.63$.
These values of $Q^*$ and $\nu$ are close to those of the 3d Ising
universality class 
0.623 \  \cite{blote:1995} and 0.630 \ \cite{guida}, respectively.

The variation of $Q_L$ as a function of $\beta^*$ for the different
volumes is shown in Fig.\ \ref{fig4}. Although there is considerable
spread in the intersection points due to correction-to-scaling
contributions, the corresponding values of $Q^*$ are close to the Ising
value (0.623) and beyond doubt permit to rule out mean field behavior 
($Q^*$ = 0.457) \cite{brezin:1985}.
Further support for Ising-like exponents is provided by the scaling of 
$< {\delta \cal M}^2 > $ at $T_c^*(L)$ versus  $L^{2 \beta/\nu}$
\ \cite{binder:1981} yielding 
$\beta/\nu = 0.52$ in accord with the 3d Ising value (0.517) and in 
clear contrast with the classical value 1 (cf.\ Fig.\ \ref{fig5}).

The ordering operator distribution $p_L(\cal M)$ at  $T_c^*(\infty)=0.04917$ is shown
in Fig.\ \ref{fig6} for the different volumes considered. 
Due to the higher value of the critical temperature  compared to that
estimated in Ref.\  \onlinecite{cai2} (0.0488) the $p_L(\cal M)$ are much closer to the
infinite system limit than those of Ref.\ \onlinecite{cai2}.

Extrapolation of the apparent chemical
potentials defined as
$\mu^*_c(L) \equiv \mu^*(\beta^*_c(L),L)$ using  a relation similar to 
Eq.\ \ref{scalt}, yields the infinite volume critical
chemical potential  $\mu^*_c=-13.600 \pm 0.005$.
Finally,
an apparent critical density $\rho^*_c(L)$ was obtained from  $\int
d\rho \ \rho p_L(\rho)$ calculated at $\beta^*_c(L)$ and $\mu^*_c(L)$.
As already remarked in Refs.\ \onlinecite{cai2} and \onlinecite{orkou:1999}
the results are nearly constant within statistical error
extrapolating to the infinite volume critical density $\rho^*_c=0.080
\pm 0.005$. The critical density remains thus unchanged from our
previous estimate \cite{cai2}.
 
The scaled distributions $p_L^{\rho}$ associated with those of $p_L(\cal M)$
obtained at 
  $T_c^*(\infty)$ (cf.\  Fig.\ \ref{fig3}) are shown in Fig.\
\ref{fig7}. With increasing system size a net tendency manifests for a
more symmetric curve with equal peak heights as expected in the limit   
$L \rightarrow \infty$. However, an increase of the statistical error with
volume is also apparent as well as the inedequate sampling at low density.

In summary, reanalysis, in the framework of the scheme of Bruce and
Wilding \cite{bruce:1992,wil1}, of new simulation results involving four
times larger volumes than considered in previous work,
increased statistics and use of  a recent determination of the order
parameter distribution of the 3d Ising universality class \cite{tsyp} leads to
i) a change of critical temperature of 0.5 \%.  
ii) an estimate of the critical exponents $\nu$ and
$\beta/\nu$ and the parameter
$Q^*$ based on the sole knowledge of the critical
temperature and parameter $\theta$ in contrast with our previous
results which were shown to be only compatible
with the Ising universality class.
These new data confirm our previous conclusion of the agreement of the critical
behavior of the RPM with that of the 3d Ising system.

In order to avoid an {\em a priori} assumption of the universality class,
Orkoulas {\em et al} \ \cite{orkou2:2001}  propose to study the scaling
properties of moments or combination of moments, $Y_j(\rho,T;L)$, of
the distribution $p_L(\rho,u)$ as, for instance, the specific heat or the
susceptibility 
$Y_{7} = \displaystyle \frac{1}{V} [< O^2> - <|O|>^2]$ 
($O = N -<N>$), computed as a
function of temperature
along an appropriate  locus in the ($T,\rho$) plane. 
Extrapolation to the thermodynamic limit of the effective
temperatures associated with the peak positions in $Y_j$ for each system
size provides estimates of
$T_c(\infty)$ and critical exponents.
We have applied this analysis to the RPM for all  the functions
displayed in Ref.\   \onlinecite{orkou2:2001} choosing
as locus  the line of inflection points of the density
versus chemical potential along an isotherm (the $\chi_{NNN}=<O^3>/V=0$ locus of Ref.\
\onlinecite{orkou2:2001}).  
The variation of $T^*_c(L)$ associated with the peak positions of the
functions $C_V$, $Y_3$, $Y_7$, $Y_{8-}$, $Y_{8+}$ and $Y_{12}$, defined in
Ref.\ \onlinecite{orkou2:2001}, along the locus $\chi_{NNN}=0$ 
is shown in Fig.\ \ref{fig8}
as a function of $L^{-1/\nu}$.
Although all functions seem  to vary nearly linearly the
extrapolated critical temperatures present a rather large scatter
between 0.0493 and 0.0490. 
Only those associated with $Y_7$ and $Y_{8+}$ are compatible with the
critical temperature 0.04917 derived from the BW f.s.s. procedure.
It is worth noticing that the statistical error on the values of $Y_j$
is difficult to estimate but a 1\% value seems to be a conservative lower
bound.   
   
An alternative approach we propose is to search for a
remarkable point (saddle point or extremum) of $Y_j$ in the whole
$(\rho,T)$ plane and measure its height as a function of volume.
Provided such a remarkable point exists and is located in the estimated
critical 
region, the height of $Y_j$ should scale as $L^{\omega/\nu}$
where $\omega$ is the exponent of the power-law type divergence of $Y_j$
in the thermodynamic limit.

All functions $Y_j$ depending explicitly on the absolute value of $O$
considered in Ref.\
\onlinecite{orkou2:2001} (as, for instance, $Y_3$, $Y_7$, $Y_8$ ...) were found to
exhibit remarkable points in the critical region. As an example we
considered the function $Y_7$ which gave the best estimate for $T^*_c$
(cf.\ Fig.\ \ref{fig8}). Figure \ref{fig9}
shows  the saddle point
present in $Y_7$.
A linear fit of the logarithm of the value of $Y_7$ at the saddle point
 versus $\ln L$, shown in Fig.\ \ref{fig10}, yields $\gamma / \nu =
1.89 \pm 0.03$.
We stress that the mentioned error is the error on the slope  inferred
from the linear regression; this error should not be assimilated with
the statistical error on   $\gamma / \nu$ which results from the error
 on the estimates of the histograms used to calculate  $Y_7$ and is
beyond reach. The value of $\gamma / \nu$ found is notably lower than
the 3d Ising value (1.967) \ \cite{guida} 
or the mean field value (2.0). This rather surprising result can be
considered as significant only when a reliable  estimate of the
statistical error on $\gamma / \nu$ is available.

On the other hand functions involving $O$  show remarkable points the
value of which should scale as $L^{n \beta/\nu}$. Unfortunately, no
sufficiently precise numerical location of these points could be achieved
 and therefore they could not be used to estimate $\beta/\nu$.

Finally, the specific heat at constant volume  $C_V/V$, calculated along the
locus \mbox{$\chi_{NNN}=0$ \ \cite{orkou2:2001}} is shown in Fig.\
\ref{fig11}. Although the peak positions shift correctly towards the
critical temperature determined above and the widths of the curves narrow with
increasing system size, there is no detectable scaling of the amplitudes
of the peaks over the volume range considered in this work. 
Similar conclusions are reached for the chemical potential term $C_{\mu}/V$ of the specific heat (cf.\ Eq.\
\ref{cv}) as evidenced in Fig.\ \ref{fig11}.
A possible explanation for the non-singular behavior of $C_V$ is that
the amplitude of the singular 
term in $C_V$  is small in the RPM  and the specific heat
dominated by 
its regular contribution,
It can also be remarked that the peak
heights in $C_V/V$ would scale, assuming
Ising value for the specific heat exponent, only by a factor
$2^{\alpha / \nu} \sim 1.12$ when doubling the linear dimensions
of the system. It is quite possible that such a small increase of peak
height is not observable within the statistical uncertainty of our
calculations.  
Results for the isochoric specific heat of the discrete lattice RPM
\cite{luij:01} show
a much more pronounced enhancement of the maximum of $C_V$ with system
size.

\section{Conclusion}
New MC simulations for system sizes up to $L/\sigma =34$, analyzed within the context of the revised
scaling theory \cite{wil1,rehr} lead to a new estimate of $T_c^*$ for the RPM, differing
by 0.5 \% 
from our earlier one \cite{cai2}, and to critical exponents $\nu$ and
$\beta/\nu$ and value of $Q^*$ in
excellent  agreement
with those of the 3d Ising universality class. This estimate relies on matching the
order parameter distribution of the Blume-Capel model obtained recently
in lattice simulations by Tsypin and Bl\"ote \cite{tsyp}.
 An analysis based on the moments of $Y_j$ which makes no 
assumption of the universality class failed to give a precise estimate
of $T_c^*$.
Furthermore, the value of $\gamma/\nu$ estimated from the critical
behavior of $Y_7$ does neither match the Ising nor the mean field value.
 The behavior of
the constant volume specific heat gives no indication of the expected
$L^{\alpha/\nu}$ scaling within the range of system sizes considered. 
In addition, the contribution $C_{\mu}$ to the specific heat does not show any anomaly which would
challenge the use of  Eqs.\ \ref{a1} and \ref{a2} for the
scaling fields \cite{fish:2000}. 
Recent studies of a discrete version of the RPM, with lattice sizes up to
$L=22$, based on a finite size scaling analysis similar to the one
considered in this work \cite{pana:02} or on the methodology
proposed in Ref.\ \onlinecite{orkou2:2001} \ \cite{luij:02} also conclude to
Ising criticality of the RPM.

\acknowledgments
%
We thank the referee for valuable comments on the manuscript. 
Computing time on the NEC SX-5 was granted by the Institut de
D\'eveloppement et de Ressources en Informatique (IDRIS).
\newpage
%

\newpage
\begin{table}
\caption{The table summarizes, for each value of the reduced simulation volume $V/\sigma^3$,
the range of temperatures $T^*$ and chemical potentials $\mu^*$ at which
simulations have been performed. $n_T$ denotes the number of different
temperatures considered  and $n_s$ the total number of thermodynamic
states.  The last two columns give the number $n_r$ of 
configurations generated in each simulation run and $n_c$ the total
number  of selected  configurations (spaced by 250 trial moves) used to compute a reweighted
histogram at given $V$.}
\begin{tabular}{|c|c|c|c|c|c|c|}
\hline
 $ V \,/ \sigma^3 $ & $ T^* $ & \ $ n_T$  &  $ n_s $&  
$ \mu^*  $ & $ n_r $  & $ n_c $  \\
\hline
 5000 &0.04928 & 1 & 1 & -13.569  &
 $ 23 \, 10^9 $&  $ 92 \, 10^6 $ \\
 10000 &\ \ 0.04915 & 1 & 2 & -13.603 to -13.605 &
 $ 15 \, 10^9 $&  $ 120 \, 10^6$ \\
 20000 &\ \ 0.0489 -- 0.04920& 4 & 4 & -13.59 to -13.65 &
 $ 87 \, 10^8 $&  $ 153 \, 10^6$ \\
 40000 &\ \ 0.04910 -- 0.04931 & 6 & 14 & -13.56 to -13.62 &
 $ 62 \, 10^8 $&  $ 390 \, 10^6$\\ 
\hline
\end{tabular}
\end{table}

\newpage

\begin{figure}
\protect \caption{ $\chi_L^2$ \  (Eq.\ \ref{xi}) as a function of $s$ for
the four volumes considered. For each volume, five temperatures are
shown differing by 0.00001. The lowest temperatures 
are 0.04920 ($V/\sigma^3$=40000),
0.04919 ($V/\sigma^3$=20000),
0.04923 ($V/\sigma^3$=10000),
0.04933 ($V/\sigma^3$=5000) and the corresponding curves are marked by
circles. Those corresponding to temperatures successively increased by
0.00001 are
marked as squares, diamonds, up triangles and left triangles, respectively.  
}
\label{fig1}
\end{figure}

\begin{figure}
\caption{ 
The apparent critical temperature $T^*_c(L)$ as a
function of $L^{-(\theta +1)/ \nu}$, with $\theta = 0.53$ and $\nu =
0.630$ obtained by matching the universal distribution calculated for
the Blume-Capel model \cite{tsyp} (circles) and the 3d Ising model
\cite{tsyp} (squares). 
In the former case the apparent temperatures are
0.04934, 0.04926, 0.04921 and 0.04922 for volumes $V/\sigma^3$=5000,
10000, 20000 and 40000, respectively,  extrapolating by linear least
square fit to the infinite volume
temperature estimate $T^*_c = 0.04917 \pm 0.00002$. 
In the latter case the apparent temperatures are
0.04932, 0.04923, 0.04920 and 0.04921 for volumes $V/\sigma^3$=5000,
10000, 20000 and 40000, respectively,  extrapolating by linear least
square fit to $T^*_c = 0.04916 \pm 0.00002$. 
L is in units of $\sigma$.}
\label{fig2}
\end{figure}

\begin{figure}
\protect \caption{
Collapse of the ordering operator distribution
function $p_L(\cal M)$ onto the universal Ising ordering operator
distribution  ${p}^{*}_{is}(x)$ for $V/\sigma^3 =
5000$, $T^*_c(L)=0.04934$ ($s=-1.45$) , $V/\sigma^3 =
10000$, $T^*_c(L)=0.04926$ ($s=-1.465$), 
$V/\sigma^3 = 20000$, $T^*_c(L)=0.04921$ ($s=-1.47$) 
and $V/\sigma^3 = 40000$, $T^*_c(L)=0.04922$ ($s=-1.43$).
The universal distribution
${p}^{*}_{is}(x)$ (solid circles) is the MC result of Tsypin and
Bl\"ote  \protect \cite{tsyp} obtained from the Blume-Capel model. The scaling variable is $x=a^{-1}_{\cal
M}L^{\beta / \nu}({\cal M} - <{\cal M}>_c)$. Scale factors
are chosen such that the distributions have unit variance.}
\label{fig3}
\end{figure}

\begin{figure}
\protect \caption{Variation of the ratio $Q_L=
{{<\delta {\cal M}^2>}_L}^2/ {<\delta {\cal M}^4>}_L$
as a function of reduced inverse temperature $\beta^*$ for the different
volumes considered. The size of the symbols is slightly smaller than the
estimated uncertainties. From top to bottom, $V/\sigma^3$=40000, 20000, 10000
and 5000, respectively. The symbols denote the simulation results and
the lines the fit by means of Eq.\ \ref{b25}.} 
\label{fig4}
\end{figure}

\begin{figure}
\caption{Variation of $\ln{{<\delta {\cal M}^2>}_L}$ at $T^{*}_c(L)$
as a function of $\ln L$. Error bars are of the order of 1 \%. The slope
of the linear least square fit is $2 \beta/\nu \approx 1.04$.} 
\label{fig5}
\end{figure}

\begin{figure}
\protect \caption{ 
 Ordering operator distribution
functions $p_L(x)$ at $T^*_c(\infty)=0.04917$ and  chemical potential $\mu^*$ determined such that $p_L(\cal M)$ is symmetric, 
 for $V/\sigma^3 =
5000$ (open squares), $10000$ (open circles), $20000$ (solid squares)
 and $40000$ (solid circles).
The universal distribution
${p}^{*}_{is}(x)$ (solid line) is the MC result of Tsypin and
Bl\"ote \protect \cite{tsyp}.
 The scaling variable is $x=a^{-1}_{\cal
M}L^{\beta / \nu}({\cal M} - <{\cal M}>_c)$. Scale factors
are chosen such that the distributions have unit variance.}
\label{fig6}
\end{figure}

\begin{figure}
\protect \caption{ 
Density distribution  $p_L^{\rho}$ at the critical temperature $T^*_c =
0.04917$ and  chemical potential $\mu^*$ determined such that $p_L(\cal
M)$ is symmetric for the volumes $V/\sigma^3$ =
$5000$ (open squares), $10000$ (open circles), $20000$ (solid squares)
 and $40000$ (solid circles).
The universal distribution
${p}^{*}_{is}(x)$ (dashed line) is the MC result of Tsypin and
Bl\"ote \protect \cite{tsyp}.
 The scaling variable is $x=a^{-1}_{\rho 
}L^{\beta / \nu}({\rho} - <{\rho}>)$. Scale factors
are chosen such that the distributions have unit variance.}
\label{fig7}
\end{figure}

\begin{figure}
\caption{Variation of $T^*_c(L)$ associated with the peak
 positions of the
 functions $C_V$, $Y_3$, $Y_7$, $Y_{8-}$, $Y_{8+}$ and
 $Y_{12}$, defined in Ref.\ 
 \protect\onlinecite{orkou2:2001} 
along the locus $\chi_{NNN}=0$  as a function
 of  $L^{-1/\nu}$.
}
\label{fig8}
\end{figure}

\begin{figure}
\caption{ 
The function   $Y_7 = \displaystyle \frac{1}{V}[<O^2>-<|O|>^2] $ with
$O=N-<N>$ in the $(T, \rho)$-plane at volume
$V/\sigma^3=5000$.  The
function is obtained from histogram reweighting using data given in
Table I.}
\label{fig9}
\end{figure}

\begin{figure}
\caption{ 
Variation of the logarithm of the  peak height of the saddle point in $Y_7$
as a function of $\ln L$.
}
\label{fig10}
\end{figure}

\begin{figure}
\protect \caption{ 
Variation of the total specific heat at constant volume $C_V/V$ and the
contribution from the chemical potential, $C_{\mu}/V$, with
temperature along the locus $\chi_{NNN}=0$ at volumes
$V/\sigma^3$=5000, 10000, 20000 and 40000.}
\label{fig11}
\end{figure}

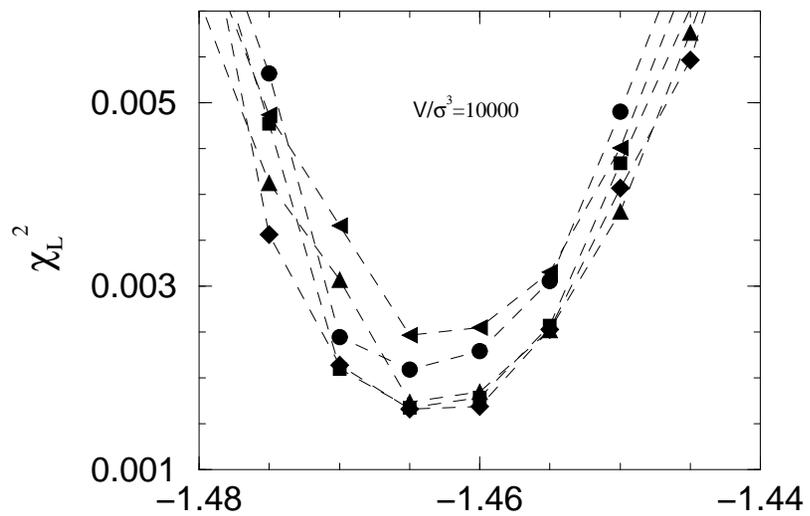
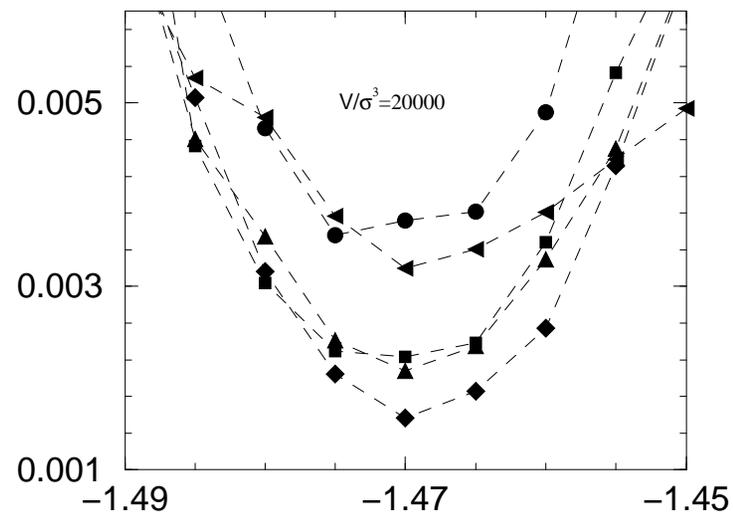
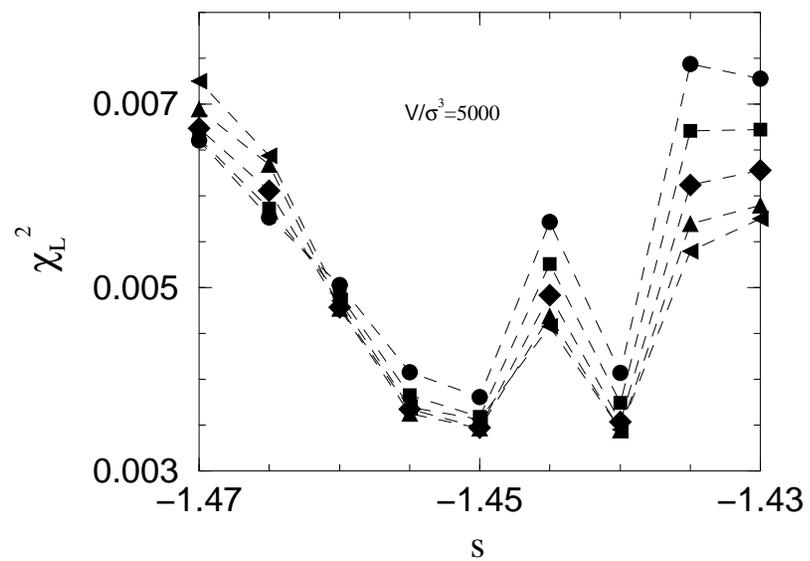
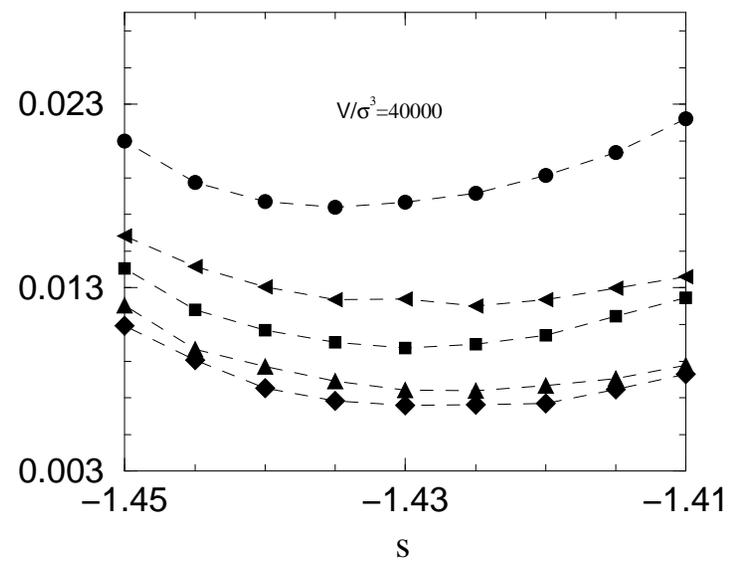
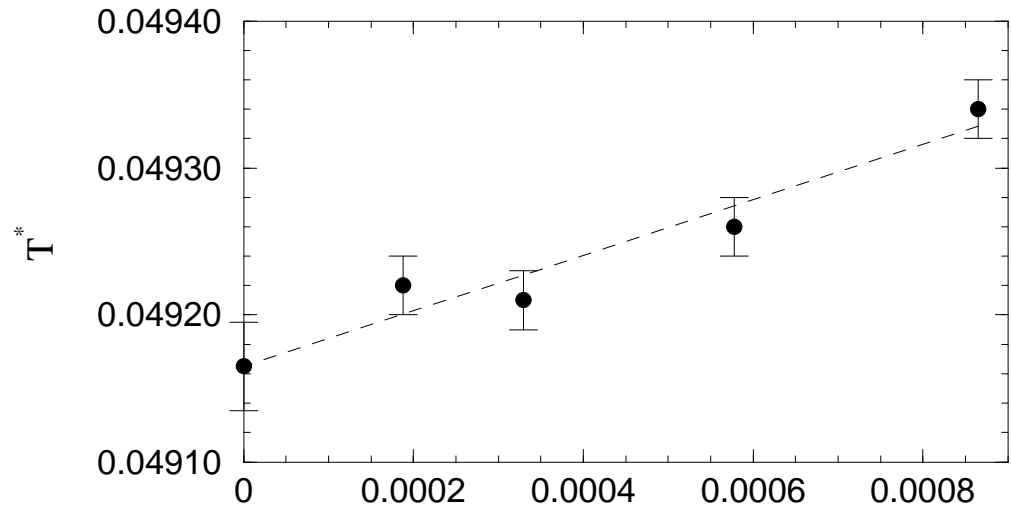
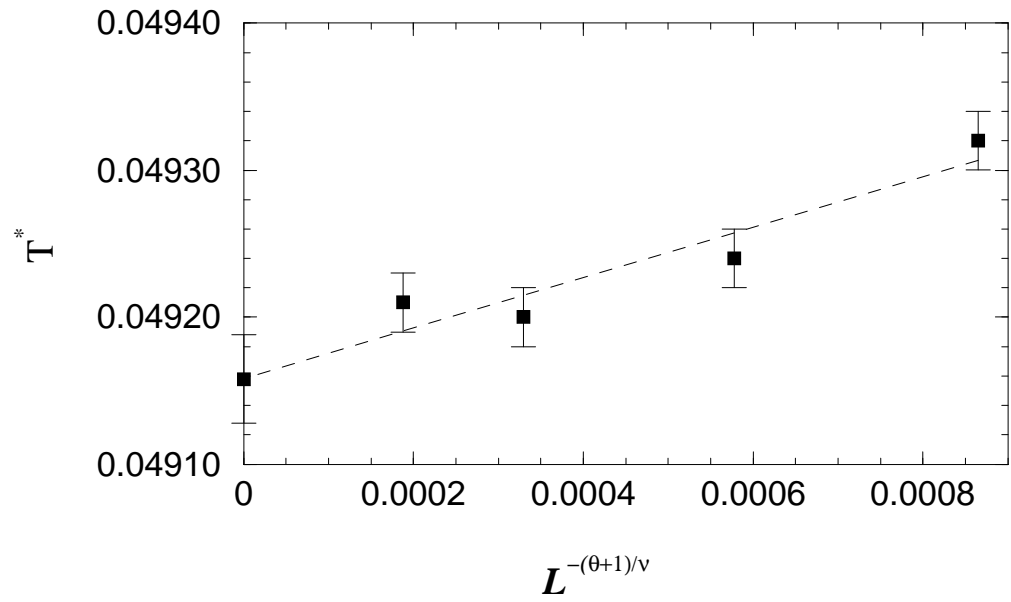
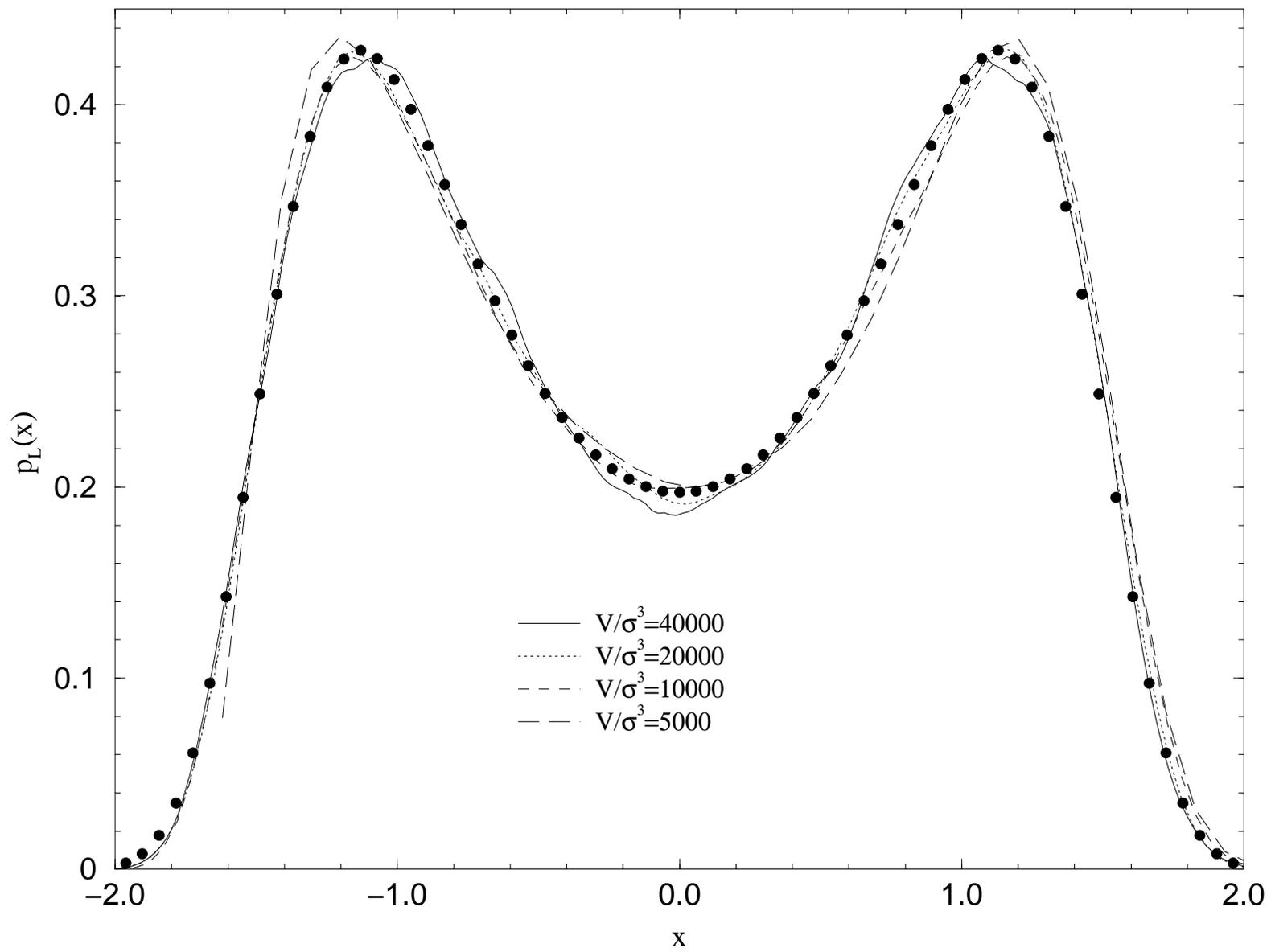
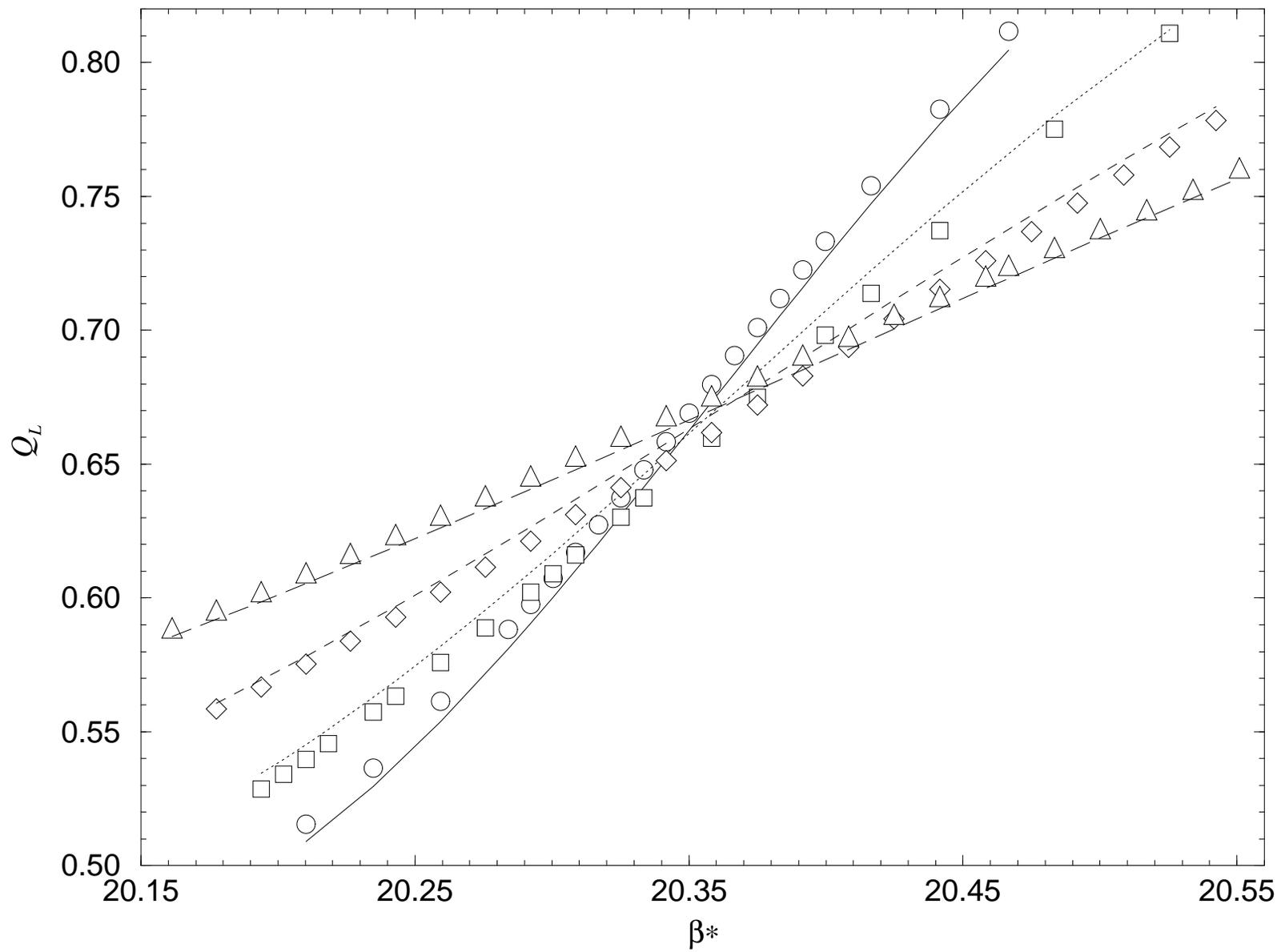
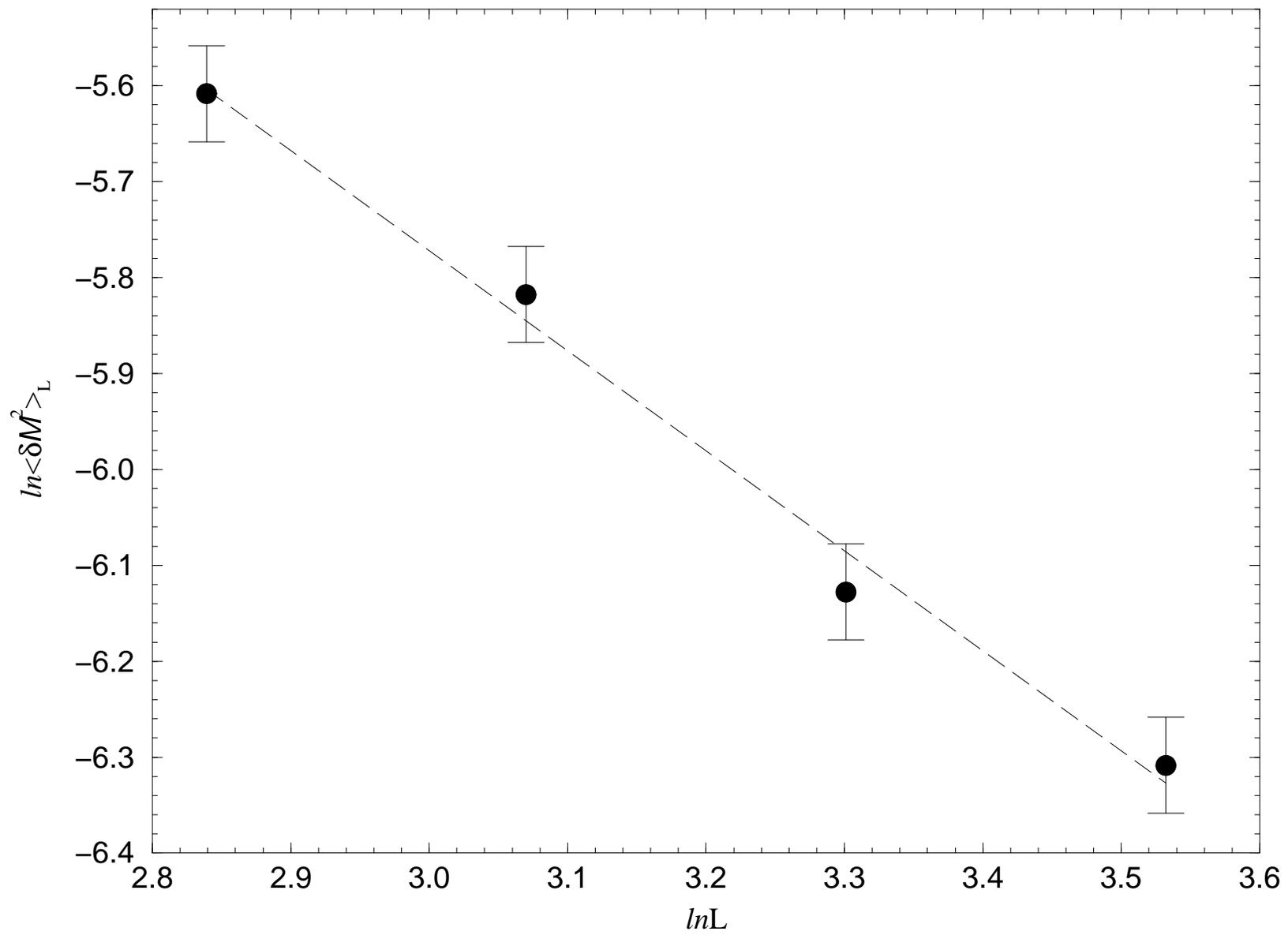
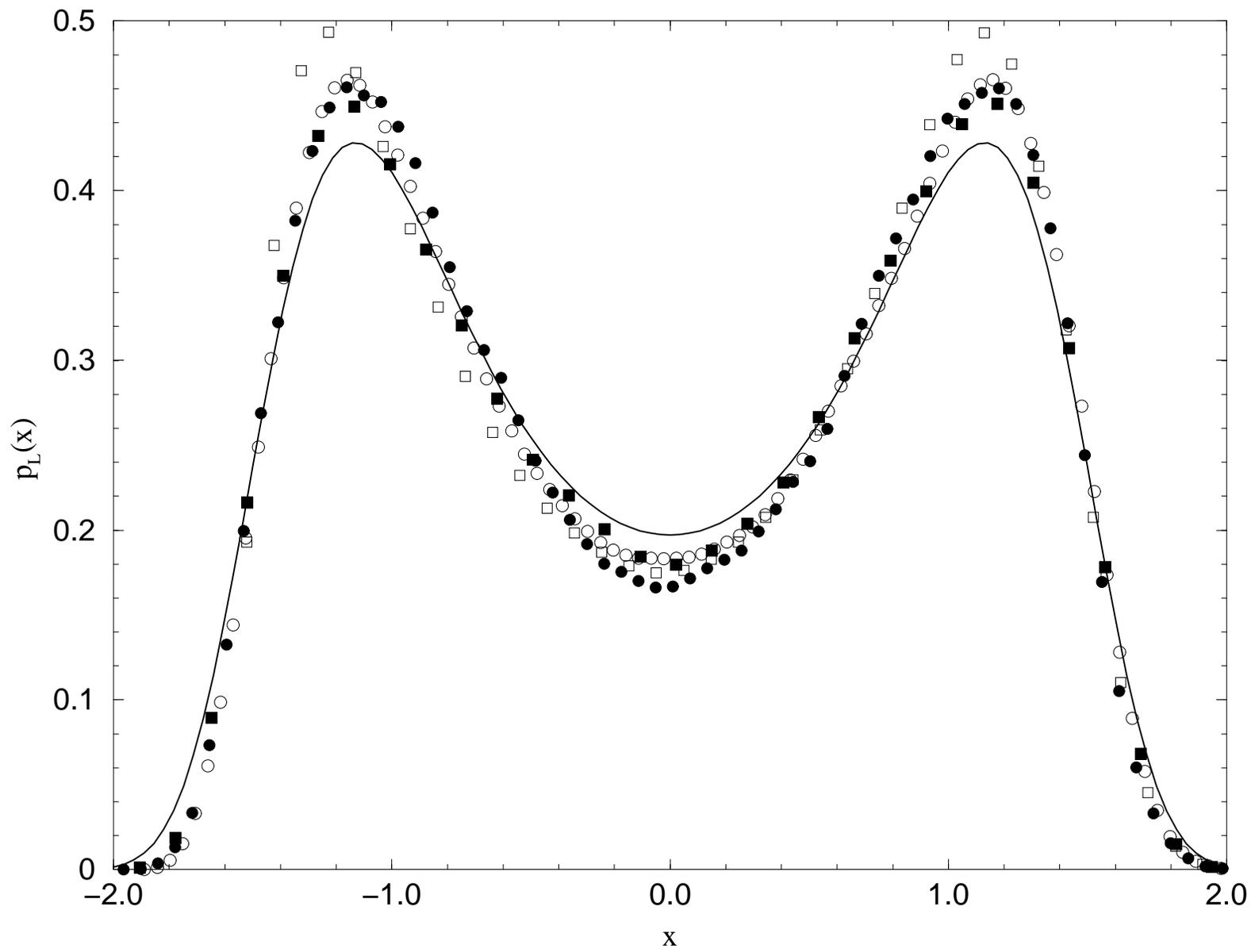
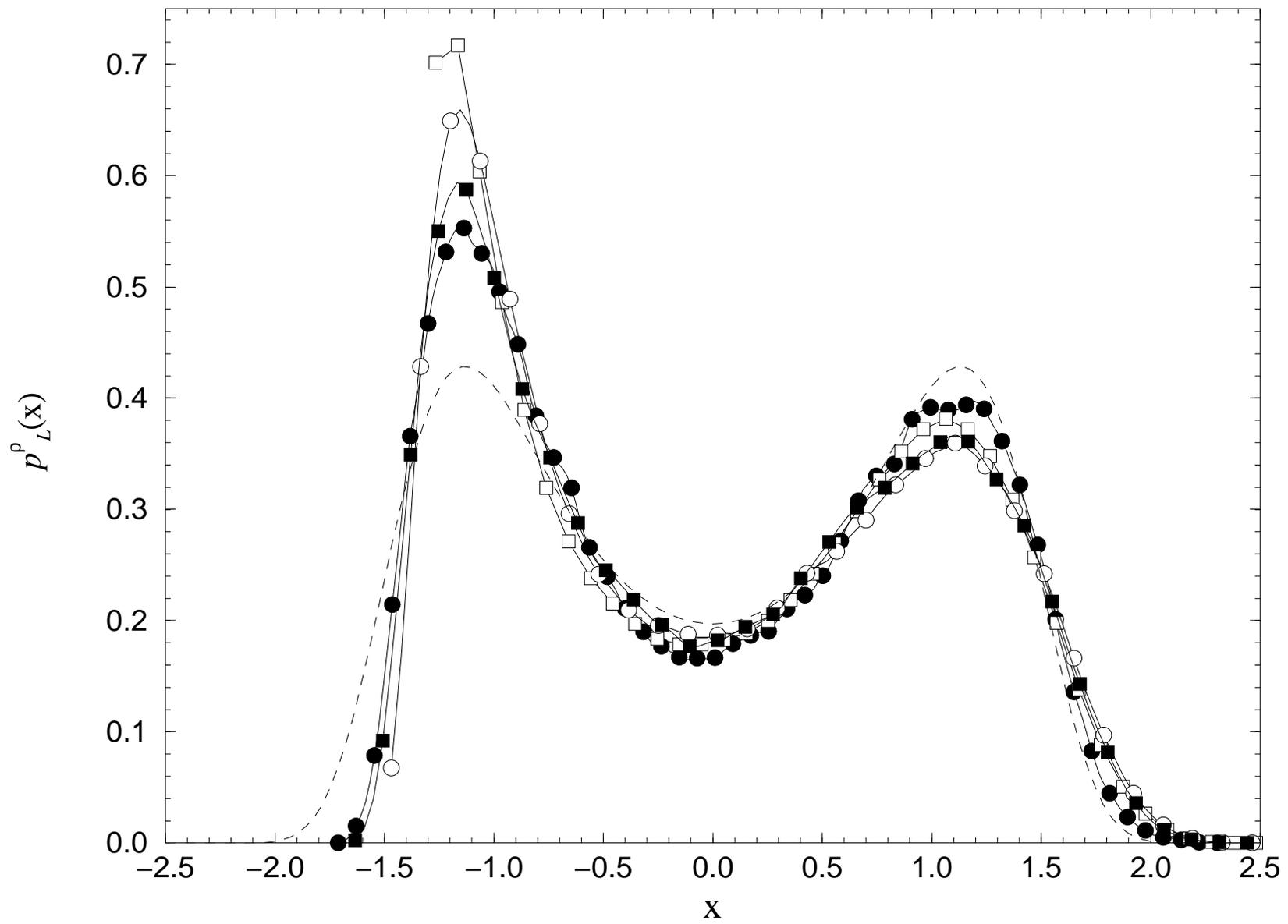
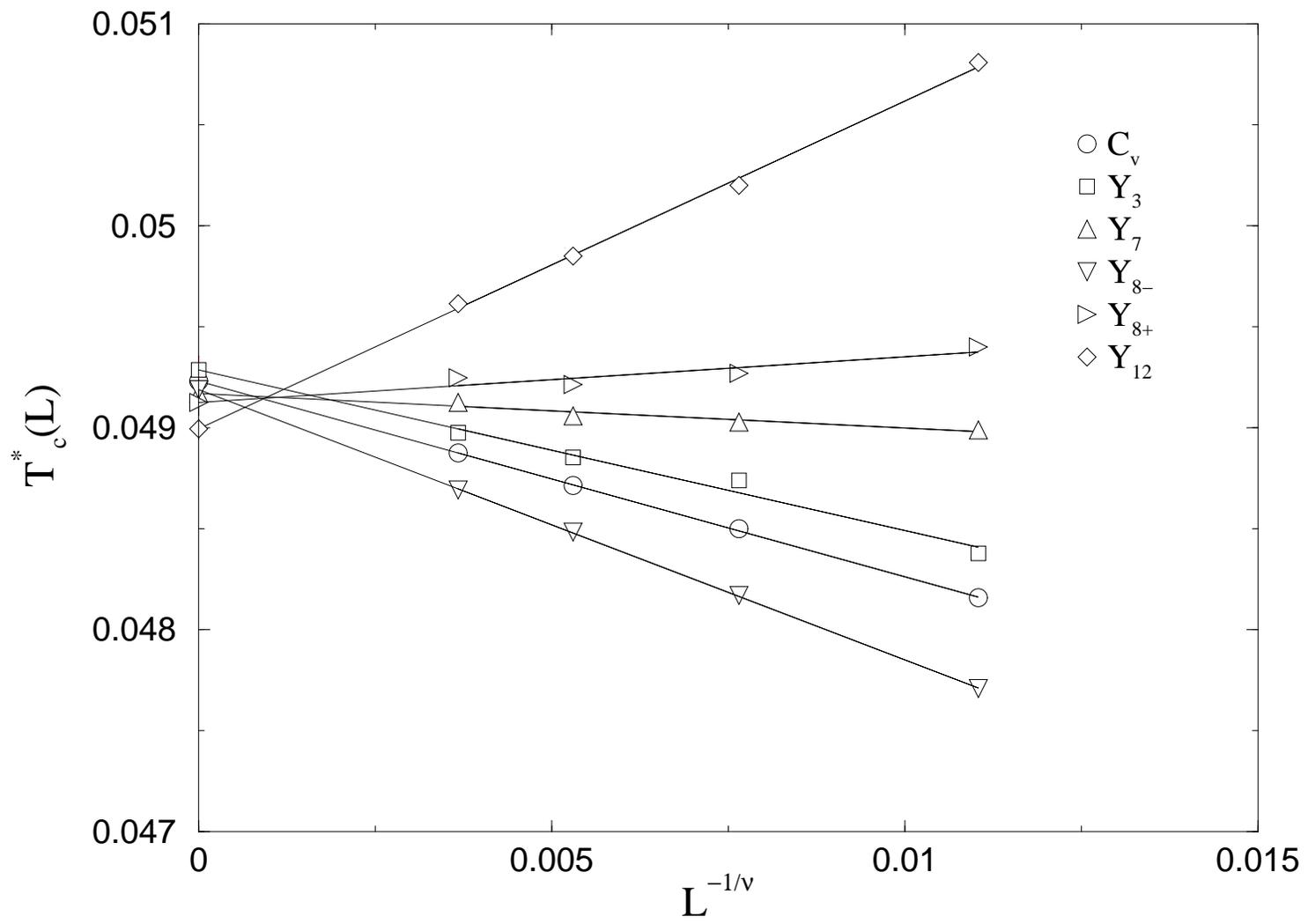
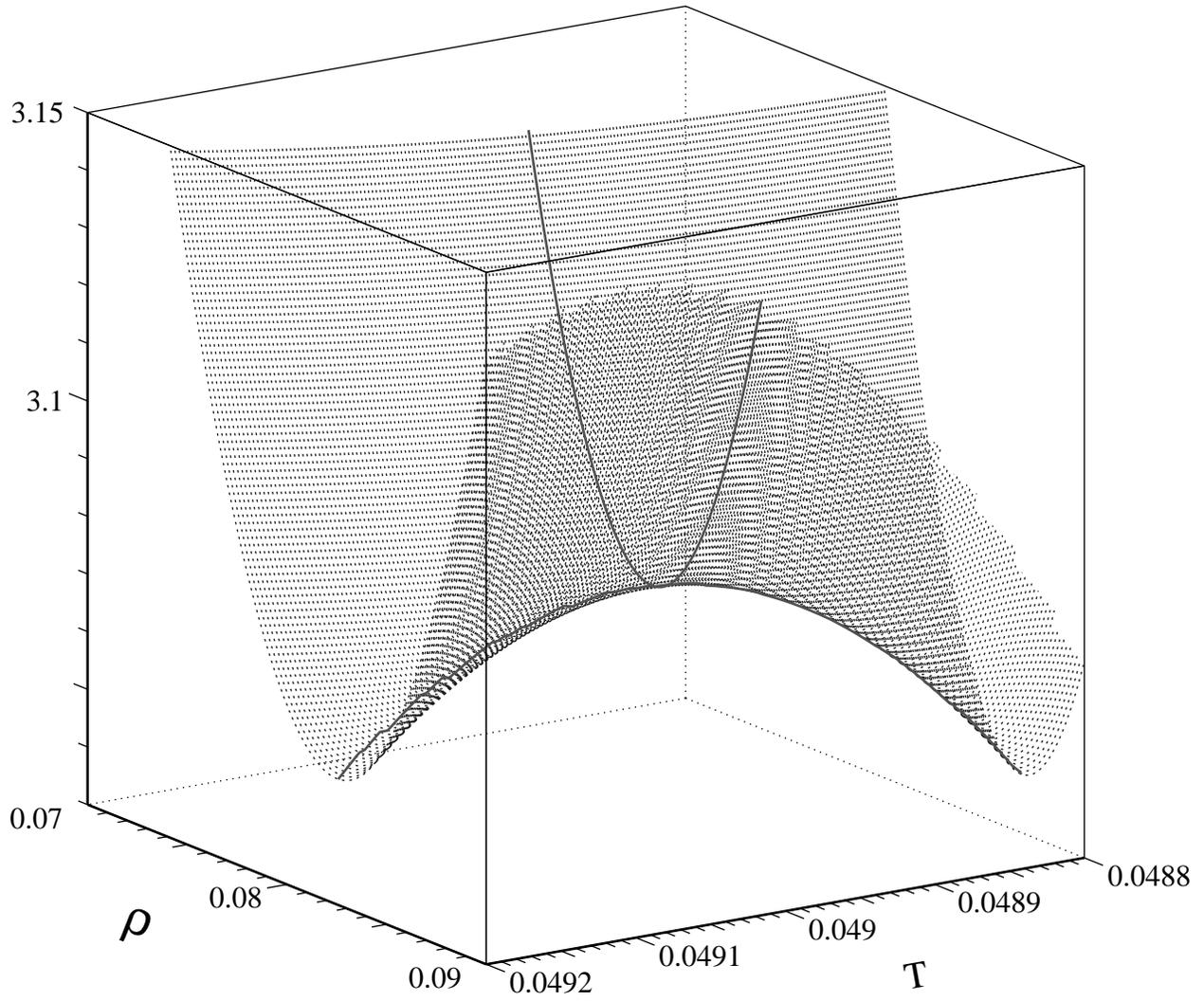
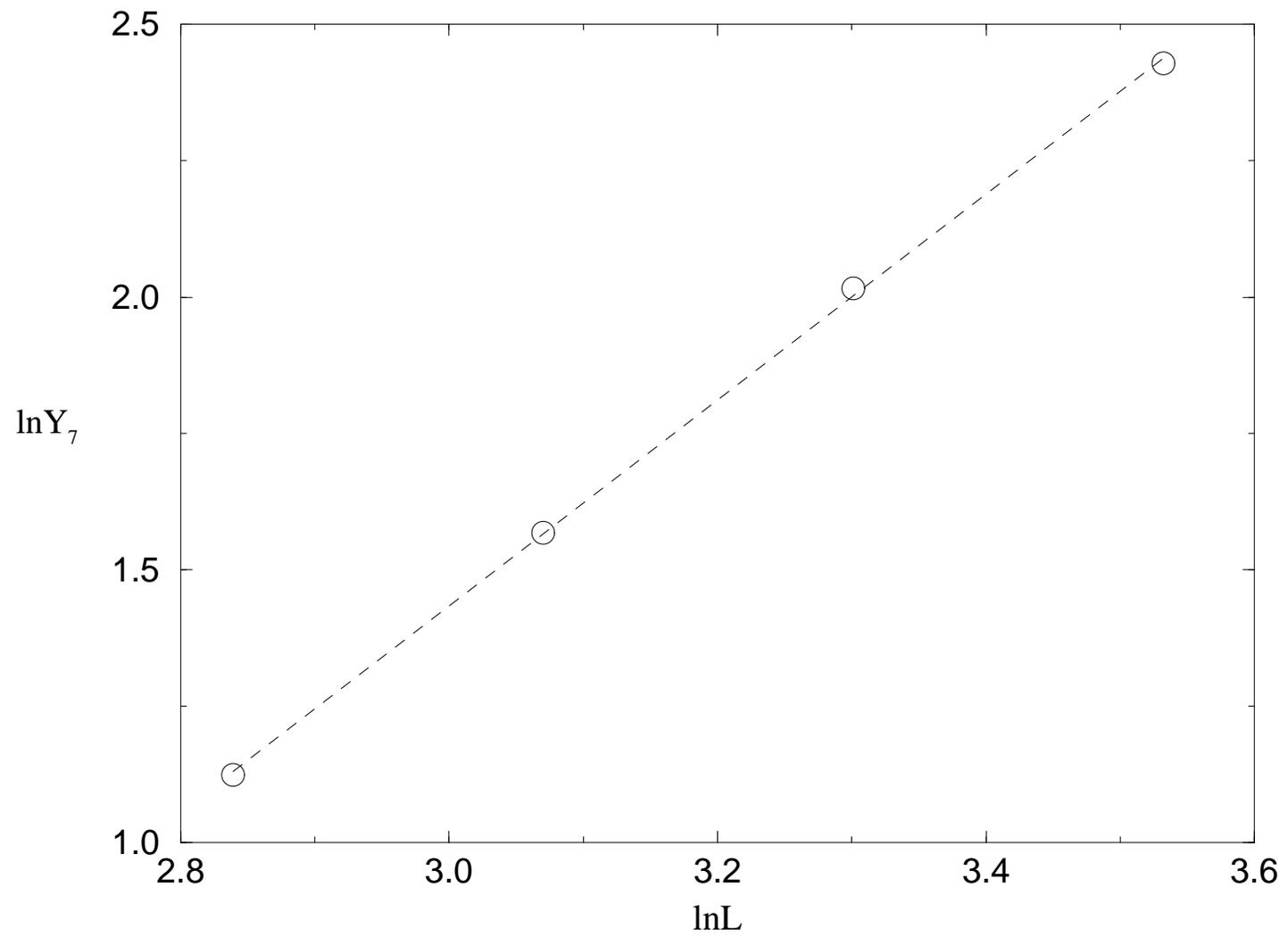
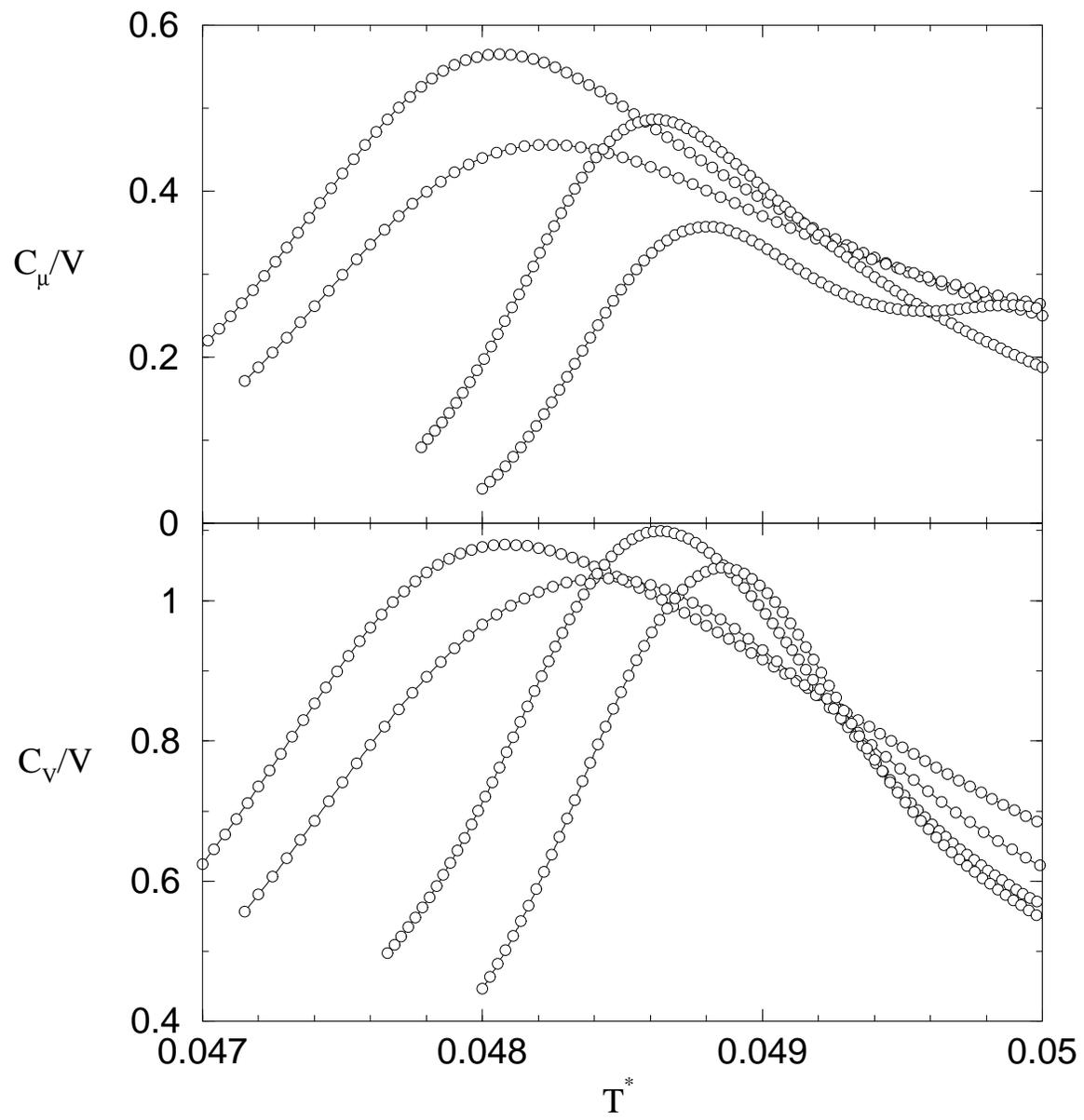
\end{document}